\documentstyle[editedvolume]{crckapb} 

\begin{opening}
\title{STAR CLUSTER SIMULATIONS}
\author{D.C. HEGGIE}
\institute{University of Edinburgh,\\
	Department of Mathematics and Statistics,\\
	King's Buildings,\\
	Edinburgh EH9 3JZ, UK}

\end{opening}
\runningtitle{STAR CLUSTER SIMULATIONS}
\def\gtorder{\mathrel{\raise.3ex\hbox{$>$}\mkern-14mu
             \lower0.6ex\hbox{$\sim$}}}
\def\ltorder{\mathrel{\raise.3ex\hbox{$<$}\mkern-14mu
             \lower0.6ex\hbox{$\sim$}}}

\begin{document}
\begin{abstract}

The simulation of rich star clusters presents challenging problems of
several kinds, including the design of suitable hardware and software,
and numerous theoretical problems in stellar dynamics and stellar
physics.  Great progress has been made possible in recent years through
the widespread use of GRAPE hardware.  Simulations are, however, still
too small to be applied to real star clusters without scaling.  How this
is done is partly an issue of stellar dynamics, and it has
thrown into focus a number of fundamental theoretical problems in this
field.

\end{abstract}

\section{Introduction}

The study of rich star clusters, each containing of order a million
stars, is a problem of stellar dynamics with ``an appealing simplicity"
(Spitzer 1987).  For many purposes the stars may be treated as point
masses, and the cluster as a many-body problem of Newtonian mechanics
under attractive inverse square law forces.  Though simply stated, this
problem is a formidable one, theoretically and computationally.

The theoretical and observational framework for understanding this
problem is summarised in the time scales quoted in Table 1.  The time
taken for a typical star to complete one orbit within the cluster is of
the order of the \emph{crossing time}.  This 
is usually defined as $t_{cr} = 2R/v$, where $R$ is a
certain measure of the radius of the cluster and $v$ is the rms speed of
the stars.  This motion is relatively fast, but the \emph{orbit} of a star
inside the cluster evolves on the much longer \emph{relaxation} time
scale, just as the orbits of the planets evolve on a much longer time
scale than the corresponding orbital periods.  Because the clusters have
lived for many relaxation times, simulations must accurately incorporate
the mechanism of relaxation.

In star clusters relaxation results mainly from the numerous mild deflections
caused by passing stars.  For a system with $N=10^4$ stars (which is the
size of a typical simulation at present), 25\% of this effect comes from
encounters with stars whose individual force contributes less than 1\%
of the total experienced by a typical star.  This is roughly the level
of error in the force calculation if this is performed by a rapid
method, such as a tree code or a particle-mesh method, as often used in
other branches of computational astrophysics.  For this reason and
others, simulations of star clusters usually avoid such methods in
favour of direct summation.

\begin{table}[htb]
\begin{center}
\caption{Time scales}
\begin{tabular}{lcc}
\hline
Time scale & Observational value & Theoretical expression\\
\hline
Orbital (crossing) time	&$\sim 10^6$yr		&$t_{cr}$	\\
Relaxation time		&$\sim 10^9$yr		&$\sim N
t_{cr}/(10\ln\gamma N)^\ast$	\\
Age$^\dagger$			&$\sim 10^{10}$yr	&---\\
\hline
\end{tabular}
\end{center}
$^\ast$ The constant $\gamma$ is discussed in Sec.3.

$^\dagger$ Throughout most of this paper the clusters we have in mind
are the old globular clusters of the Galaxy.

\end{table}

\bigskip\bigskip

The familiar problem with direct summation is that it is very
time-consuming.  The effort in a simulation grows with $N$ roughly as
$N^3$, where two powers stem from the force calculation and the third
reflects approximately the $N$-dependence of the relaxation time (cf.
Table 1).  The result is that, while the speed of computers has grown by
many orders of magnitude since 1960 (when the first simulations
were published), the largest value of $N$ has grown
by only an order of magnitude per decade (Fig.1).

\begin{figure}
\includegraphics{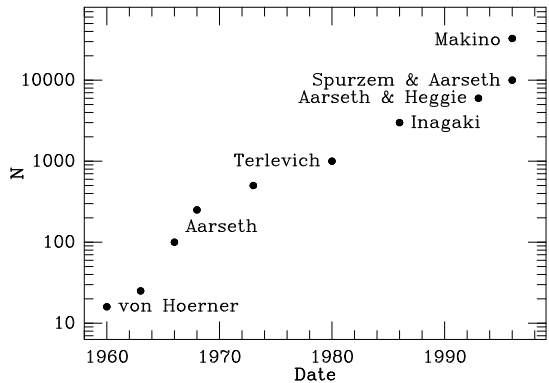}
\vspace{6cm}  
\caption{The number of stars in the largest advanced cluster simulation
as a function of publication date, with authors' names.}
\end{figure}

Evidently the largest simulations still fall far short of requirements: 
$N$ must rise by at least another order of magnitude before direct simulations
of even a modest star cluster become possible.  For this reason
considerable effort is now under way to understand how best to scale the
results of simulations to the evolution of a real cluster.  This
research has raised some intriguing and fundamental questions, which
will be discussed in Section 3 of this paper.  First, however, we turn
to some of the hardware issues with which such simulations confront us.

%
%
%
%

\section{Hardware Problems}

Of the last three points in Fig.1, one (Aarseth \& Heggie) results from
use of a workstation, one (Spurzem \& Aarseth) from a supercomputer and
the third (Makino) from a special-purpose computer.  This suggests that
all three are roughly competitive in the simulation of large star
clusters.  The supercomputer calculation, however, took two months,
spread over a period of two years (Spurzem \& Aarseth 1996), and there
would be no question of conducting repeated runs, as would be required
in most useful scientific investigations.  The workstation run took some
months (Aarseth \& Heggie 1993), and again it would be necessary to
reduce $N$ considerably for routine work.  The run on a special-purpose
computer took about 3 months (Makino 1996), but what is important is
that it yielded by far the largest $N$.  With this hardware it is
possible to conduct large numbers of slightly more modest simulations, with
$N\sim10^4$, in a reasonable time.

It is largely for this reason that simulations on special-purpose
(GRAPE) hardware have come to dominate this subject in the last few
years.   Another very important reason is that the GRAPE team, led by
Professor Sugimoto, have actively encouraged research groups abroad to
adopt copies of the hardware.  Copies of the high-precision hardware are
now in operation in the UK, Germany and the USA.  These installations
are much smaller than the prize-winning Teraflops GRAPE at the
University of Tokyo, but in fact for simulations of the relevant size
their performance is comparable (Aarseth \& Heggie 1997).  On the
larger of the two boards at the Institute of Astronomy (Cambridge,
UK) a simulation with $N=32K$ (i.e. $N = 32768$)
covers a time corresponding to the age of
a globular star cluster in about 200 hours, and the time is roughly
proportional to $N^2$.  Note that the power of $N$ is smaller than expected
(cf. the Introduction) because the efficiency of the hardware increases
with $N$.  It should be stressed, however, that this time depends on the
complexity of the model:  systems with numerous binary stars, for
example, would take much longer.

It is interesting that not one of the points in Fig.1 comes from work
with a modern, general-purpose, fine-grained parallel supercomputer.  

\section{Scaling Problems}

\subsection{An outline of the issues}

Despite the advances of decades of development, culminating in the
present generation of GRAPE hardware, the largest useful $N$-body
simulations are still too small, by more than an order of magnitude. 
Nevertheless, efforts are being made to apply these techniques to
quantitative problems in the evolution of star clusters.  For example,
stars of low mass are preferentially driven out of a cluster as a result
of two-body encounters, and Vesperini \& Heggie (1997) have used
$N$-body models to study the resulting changes in the relative abundance
of stars of different mass.  Recent observational advances have made
this issue an important one.

The problem for simulations is that different dynamical processes
develop on time scales which depend in different ways on $N$, as is
already apparent from Table 1.  Therefore when the results of an
$N$-body simulation are scaled to a real cluster, where $N$ is much
larger, it is necessary to understand the dominant mechanism which
determines the evolution.   This can vary from one problem to another. 

Consider, for example, a cluster with low initial density and large
numbers of stars of high mass.  These stars evolve rapidly, on a time scale of
a few million years, by their internal evolution, and towards the end of
their lives they lose a great deal of mass.  This loss of mass can lead
to the rapid escape of stars, and total disruption of the cluster in
much less than $10^9$ years.  In this time little relaxation takes
place, and stars escape on an orbital time scale.  In scaling a
simulation to the evolution of such a cluster, therefore, we should
ensure that the \emph{crossing} time of the simulation scales to that of the
cluster (Fukushige \& Heggie 1995; McMillan, pers. comm.). 

In the following subsection we consider the opposite extreme, i.e.
a cluster born with neither too low a density nor too many high-mass
stars.  Indeed, for understanding the old star clusters of our galaxy
this is the more important case.  The evolution of such clusters is
dominated by relaxation, except possibly for an early phase of expansion
associated with the evolution of the stars of high mass.  For such
clusters, scaling by the \emph{relaxation} time is appropriate.

To make clear what is involved, consider the formula 
\begin{equation}
t_{rh} = 0.138{N^{1/2}r_h^{3/2}\over m^{1/2}G^{1/2}\ln\Lambda}
\end{equation}
for the \emph{half-mass relaxation time} (Spitzer 1987), in terms of the
\emph{half-mass radius} $r_h$, i.e. the radius of a sphere containing
the 
inner half of the mass of the cluster, the mean stellar mass $m$, the
constant of gravitation $G$, and the \emph{Coulomb logarithm}.  Its
argument is usually taken as 
\begin{equation}
\Lambda = \gamma N,
\end{equation}
where $\gamma$ is a constant of order unity (see below).  For a given choice of
$\gamma$ we compute $t_{rh}$ for both the $N$-body simulation and the
cluster which we intend to model.  This allows us to convert times in
the simulation to millions of years.

The following subsection examines scaling by the relaxation time in
considerable detail, but other scaling problems will require solution in
future.  One of these concerns the fact that a typical star cluster has
an elongated orbit, which results in repeated disturbances as the
cluster passes near the Galactic bulge.  The problem here is that the
period of orbital motion of the cluster scales with the crossing time.
If the cluster survives long enough that relaxation is important, as
discussed above, then we are confronted with two evolutionary mechanisms
of comparable strength operating throughout the life of the cluster on
time scales which depend in different ways on $N$.  It is not known how
to scale results of simulations in this situation.

\subsection{Scaling by the relaxation time}

An obvious test of a scaling procedure is to check that the results do
not depend on the size (i.e. the value of $N$) of the simulation, except
for statistical fluctuations.  Suitable material for such a test became
available when, in 1997, the author initiated a collaborative experiment in the
simulation of star clusters.  This experiment was not restricted to
$N$-body methods, but that is what we concentrate on here.  The results
of this experiment will be reported elsewhere, but in the meantime much
information on the project can be found at the web site\hfill\break
http://www.maths.ed.ac.uk/people/douglas/experiment.html

The initial conditions of this experiment specified the initial
distribution in position, velocity and mass of the stars, and the
circular orbit of the cluster around its parent galaxy.  No attempt was
made to include any initial population of binary stars or stellar
evolution, and so the dynamical evolution of the cluster is dominated by
two-body relaxation.  These choices were made as a compromise between an
interesting, realistic problem and the restricted capabilities of the
codes available.

The experiment resulted in large amounts of data, but here we
concentrate on one issue.
For astrophysicists one of the most interesting predictions of such
models is the \emph{lifetime} of the cluster.  As we shall see, stars
escape from a cluster rather rapidly, in the sense that their motion
relative to the cluster becomes dominated by the gravitational
attraction of the galaxy rather than the other cluster members, and they
never return (Fig.2).  As a result, it is estimated that several
clusters dissolve completely in our galaxy in each billion years (Hut
\& Djorgovski 1992, Vesperini 1997).   A knowledge of how the lifetime
depends on the initial conditions should help to explain the distribution of
those that still survive, and so we concentrate on the time scale of
mass loss by escape.

\begin{figure}
\includegraphics{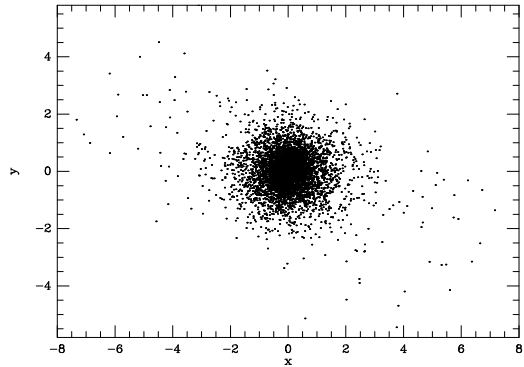}
\vspace{6cm}  
\caption{Snapshot of a star cluster simulation.  The positions of the
stars are projected onto the plane of motion of the cluster round the
galaxy.  The galactic centre lies far off along the $x$-axis.  Stars
escape most easily towards or away from the galactic centre, just as
tides are raised on the earth in directions towards and away from the
sun and moon.}
\end{figure}

For the collaborative experiment three groups submitted results from
$N$-body simulations performed with independently written codes.  All
three groups scaled their $N$-body results by the relaxation time, and
all three found the same trend with $N$, which is that the total mass
evolves more rapidly, and the cluster dissolves sooner, for larger $N$.
Table 2 presents results (in billions of years) from one of the groups.

\begin{table}[htb]
\begin{center}
\caption{Cluster lifetime (Aarseth \& Heggie)}
\begin{tabular}{lccccc}
\hline
N		&2048	&4096	&8192	&16384	&32768	\\
Lifetime (Gyr) 	&$22.6\pm0.8$	&$22.1\pm0.5$	&$20.7\pm0.4$	&
$18.6\pm0.2$	&$17.1$	\\
No. of cases$^\ast$	&4$^\dagger$&8	&4	&2	&1	\\
\hline
\end{tabular}
\end{center}
$^\ast$  Except for $N=32768$, results are averaged over several
cases.

$^\dagger$ The set for $N = 2048$ is incomplete. 
\end{table}

The results in table 2 give no information on the rate of escape at
intermediate times, and are complicated by the very issue of scaling
which we wish to test, especially by the choice of the constant
$\gamma$ in the Coulomb logarithm (eq.[2]).  Therefore more complete
results are given in Fig.3, in units in which the crossing time is
constant. 

\begin{figure}
\includegraphics{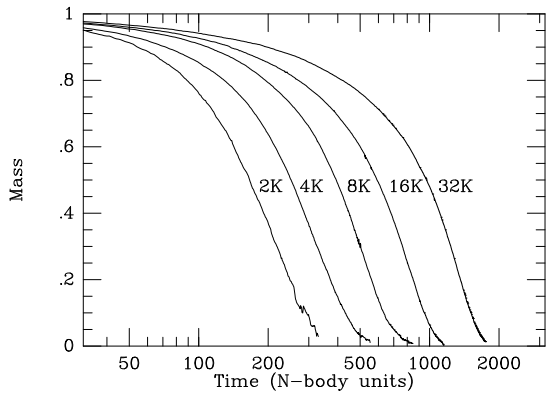}
\vspace{6cm}  
\caption{Mass as a function of time for simulations of different $N$, in
units in which the crossing time is constant.  For all except the run
with $N = 32K$, i.e. $32768$, the result is averaged over the number of
runs given in Table 2.}
\end{figure}

Now it is possible to test the requisite scaling by comparing the times
at which each set of models reaches a given mass.  Results are displayed
in Fig.4.  For example, when the mass is $0.5$ (expressed as a fraction
of its initial value) the times for the models with $N = 16K$ and $32K$
are $t = 596$ and $975$ respectively, and their ratio is $1.64$.  The curves
of Fig.4 are obtained by the same arithmetic for other values of the
mass, and other successive sets of models.  

\begin{figure}
\includegraphics{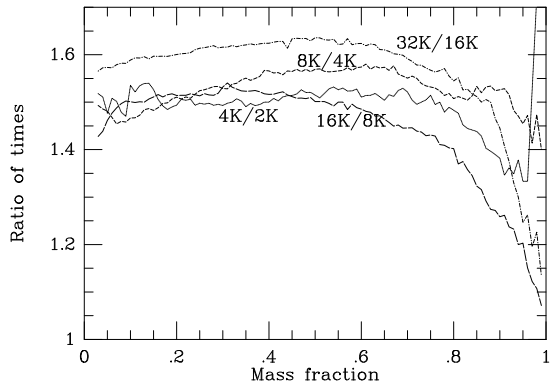}
\vspace{6cm}  
\caption{Ratios of times at which a given fraction of mass remains, for
various pairs of simulations.  If scaling by the relaxation time is
valid, the result in each case should be the ratio of relaxation times.}
\end{figure}

If scaling by the relaxation time is correct, the curves of Fig.4 should
lie at the values of $t_r(2N)/t_r(N)$, where $t_r(N)$ is a measure of
the relaxation time for a simulation with $N$ particles.  By eqs.(1) and
(2) this is $2\ln(\gamma N)/\ln(2\gamma N)$, $= 2/(1 + \ln 2/\ln(\gamma
N))$.  The value $\gamma = 0.4$ (Spitzer 1987) is often adopted, and so
for the largest runs displayed here the ratio of times should be about
$1.85$, a value clearly contradicted by the data of Fig.4.  This is the
essence of the scaling problem.

\subsection{Resolution of the scaling problem}

To a theorist with long experience in stellar dynamics the results of
Fig.4 are very perplexing, and also intriguing.  Here we present a
speculative list of possible explanations.  Objections can be devised
for every single explanation in this list, and it is possible that
several mechanisms are contributing.  Only further research can
determine which mix of explanations is correct, or whether some new idea
is needed.

\begin{enumerate} 

\item \emph{The Coulomb logarithm}:  the coefficient $\gamma$ in eq.(2)
is not rigorously determined by theory.   In order to account for the
data in Fig.4, however, a value as small as about $0.001$ would be
necessary for $N = 16K$ and about $0.004$ for $N = 2K$.  Such values are
in contradiction with any theory, any previous empirical determination,
and even the assumed $N$-dependence of $\Lambda$ in eq.(2).  A more
promising possibility is that, even for fixed $N$, the value of $\gamma$
should vary with the masses of the interacting stars, as is  suggested
by the theory of H\'enon (1975), though this would imply that no overall
scaling is feasible.

\item \emph{Large-angle scattering}:  the relaxation time is based on
the weak scattering approximation of the collision term of the Boltzmann
equation, and neglects terms which are smaller by a factor of order
$1/\ln\Lambda$ (H\'enon 1960a).  Perhaps the results of Fig.4 are
approaching the expected value, but the convergence is logarithmic.

\item \emph{The escape rate}:  the arguments that the rate of escape
scales with the relaxation time are not rigorous.  For example, long ago
H\'enon (1960b) gave a formula for the escape rate which does not include
the Coulomb logarithm.  Though this would not explain the data of Fig.4
either, the fact that the discrepancy with more conventional theories
has never been resolved indicates that the theory of escape is still
incomplete.  Several known complications of the process of escape are
not captured by existing theory.

\item \emph{New relaxation processes}: the motion of a star inside the
cluster is subject to the tidal field of the rest of the galaxy, and
this makes the motion chaotic, even in the absence of two-body encounters. 
The resulting diffusion in phase space
is not incorporated in any theory for the dynamics of star clusters. 

\item \emph{Errors}: it is well known (Miller 1964, Goodman \emph{et al} 1993)
that the detailed results on individual particles in $N$-body
simulations are wrong, and it is little more than an act of faith to
suppose that the statistical results (such as the escape rate) are
correct.  This would be an argument of last resort, however.

\end{enumerate}

\section{Conclusions}

In this paper we have described some aspects of the present status of
$N$-body simulations of large star clusters.  We have seen that progress
is currently dominated by the availability of special-purpose hardware
developed by the GRAPE group at the University of Tokyo.

Even with the power of this hardware, the size of the largest
simulations that are feasible routinely falls short of the size of the
real systems of interest in astronomy, by at least an order of
magnitude.  This shortfall imposes the need for scaling of the results
of $N$-body simulations.  The theory of relaxation is the obvious basis
for carrying out this scaling, but we have seen that the standard theory
of relaxation fails to account adequately for the consistent scaling of
simulations of different size.  Some possible reasons for this have been
listed and discussed, but the cause of the discrepancy has not yet been
clearly identified.  It is important to do so, because the resulting
uncertainty in predicting the time scale for the evolution of star
clusters is of order a factor of two.

This current dilemma illustrates two interesting features of $N$-body
simulations in this field.  First, the simulations themselves are
leading to a rather fundamental re-examination of some of the basic
theoretical ideas in the field of stellar dynamics -- ideas which have
remained as a virtually unchallenged cornerstone of the subject since
their first careful development by Chandrasekhar (1942) over 50 years ago. 
The second point of interest is the nature of the simulations.  It is
often glibly assumed that $N$-body simulations, by contrast with more
approximate models, rely less on simplifying approximations and
assumptions.  In fact, if they are to be scaled to the real systems of
interest, this needs to be done on the basis of a thorough theoretical
understanding of the mechanisms at work.  What the $N$-body models are
doing is refining this understanding, and not replacing it.

\section{Acknowledgements}  I wish to acknowledge, with gratitude and
admiration, the immense help provided to the UK community of
computational astrophysicists by Professor Sugimoto and members of his
team.  Support from the UK Particle Physics and Astronomy Research
Council under grant GR/J79461 is also acknowledged.  I am grateful to
several colleagues, including H. Baumgardt, K. Engle, T. Fukushige, P.
Hut, S.L.W. McMillan, J. Makino, S.F. Portegies Zwart and W. Sweatman,
for contributing $N$-body data to the collaborative experiment mentioned
in Sec.3.2, and/or ideas to the list in Sec.3.3.  I also warmly thank
S.J. Aarseth for the development and use of the $N$-body code used to
produce the data discussed in Sec.3.2.

\end{document}